\documentclass[prl,superscriptaddress,notitlepage,showpacs,twocolumn,floatfix]{revtex4-2}
\usepackage{float,xcolor}
\usepackage{mhchem}
\usepackage{amsmath}
\usepackage{amssymb}
\usepackage{graphicx,array}
\usepackage{txfonts}
\usepackage{mathtools}

\newcommand{\beq}{\begin{equation}}
\newcommand{\eeq}{\end{equation}}
\usepackage{color}
\usepackage{latexsym}
\usepackage{xspace}
\usepackage{ulem}

\usepackage{appendix}
\graphicspath{{figures/}}

\newlength{\fighskip} \fighskip=2pt
\newlength{\figvskip} \figvskip=3pt
\newcommand*{\figbox}[2]{{
\def\figscale{#1}
\def\arraystretch{0.8}
\arraycolsep=0pt
\begin{array}{c}
\vbox{\vskip\figscale\figvskip
\hbox{\hskip\figscale\fighskip
\includegraphics[scale=\figscale]{#2}}}
\end{array}}}
 
\begin{document}
\title{Page Time as a Transition of Information Channels: High-fidelity Information Retrieval for Radiating Black Holes}

\author{Ran Li}
\email{Corresponding author: liran@qfnu.edu.cn}
\affiliation{Department of Physics, Qufu Normal University, Qufu, Shandong 273165, China}

\author{Xuanhua Wang}
\email{Corresponding author: wangxh@ucas.ac.cn}
\affiliation{Center for Theoretical Interdisciplinary Sciences, Wenzhou Institute, University of Chinese Academy of Sciences, Wenzhou, Zhejiang 325001, China}

\author{Kun Zhang}
\affiliation{School of Physics, Northwest University, Xi'an, Shaanxi 710069, China}

\author{Jin Wang}
\email{Corresponding author: jin.wang.1@stonybrook.edu}
\affiliation{Department of Chemistry, and Department of Physics and Astronomy, The State University of New York at Stony Brook, Stony Brook, NY 11794, USA}

\begin{abstract}
The effective field theory description of a radiating black hole introduces redundant degrees of freedom that necessitate annihilation of those modes at late stages to conserve entropy. The prevailing view is that such effective process can result in information loss unless the redundant states are annihilated in maximally entangled pairs, resembling quantum teleportation. In this Letter, we demonstrate that this view can be relaxed in a new postselection model. We investigate information recoverability in a radiating black hole through the non-unitary dynamics that projects the randomly-selected modes from a scrambling unitary. We show that the model has the merit of producing the von Neumann entropy of black holes consistent with the island formula calculation and that information in the black hole interior can be decoded from the Hawking radiation without loss after the Page time. Moreover, in this model the Page time gains a new interpretation as the transition point between two channels of information transmission when sufficient amounts of effective modes are annihilated inside the horizon. We present two decoding strategies along with their quantum circuit realizations. The experimental verification of the strategies employs 7-qubit IBM quantum processors, demonstrating the viability of these strategies and the potential for quantum processors to probe the black hole interior.

\end{abstract}

\maketitle

{\it Introduction.--}
From the effective field theory description, Hawking radiations from a black hole are in a mixed state, entangled with the right-moving partner modes inside the black hole \cite{Hawking:1975vcx}. This picture runs into problems with the entropy conservation near the end of the evaporation when all the left is the mixed-state radiation. This is widely known as the information paradox \cite{Hawking:1976ra}. Though at add with information conservation, this description was for long considered valid till the very end of the evaporation. The effective field description is derived from quantum field theory in the weak gravity regime and loses information in the evaporation process, while the unitary description, which is induced from the dual boundary theory, has far fewer degrees of freedom and conserves information \cite{Page:1993df,Page:1993wv,Maldacena:1997re,Almheiri:2020cfm}. The dichotomous descriptions necessitate a reconsideration of the effective field dynamics in the black hole interior.

One early attempt to reconcile the two disparate descriptions involves the  Horowitz-Maldacena (HM) final-state projection proposal \cite{Horowitz:2003he}. It modifies the effective description by imposing a unique state at the singularity as an extra boundary condition, realized by projections of interior modes onto a highly entangled state. This mechanism allows information to escape the black hole in a manner reminiscent of quantum teleportation. Nevertheless, this model encounters challenges, including the information loss when interactions inside the black hole are considered, violation of causality, and inconsistency with the Page curve \cite{Lloyd:2013bza,Gottesman:2003up,Lloyd:2004wn,Lee:2020aft,Wang:2023eyb}. Inspired by the quantum extremal surface (QES) calculation which reproduced the Page curve, later revisions of the HM model include the island=measurement proposal on the horizon \cite{Wang:2023eyb} and the incorporation of the Haar random unitaries with the projections of the effective modes (AEHPV proposal) \cite{Akers:2022qdl,Gyongyosi:2023sue,DeWolfe:2023iuq}. In addition, the AEPHV proposal offers a new interpretation of such postselection models as a non-isometric encoding of the effective degrees of freedom into the fundamental ones. In their dynamical models, the infalling information escapes from a black hole through the EPR projections as in quantum teleportations. Whether these modifications also suffer the same ailments such as the information loss remains a crucial question to be addressed.

Motivated by the progress in black hole entanglement entropy, pivotal attention has been directed towards strategies for retrieving escaped information from radiation \cite{Yoshida:2017non,Yoshida:2018vly,Landsman:2018jpm,Brown:2019rox,Bao:2020zdo,Li:2021mnl}. The first explicit decoding strategies were provided by Yoshida and Kitaev following the milestone paper by Hayden and Preskill that reveals the rapid information escape from the black hole \cite{Hayden:2007cs,Yoshida:2017non}. On one hand, these studies jointly demonstrate the rapidity of deciphering the information fallen into a black hole from the Hawking radiation at a high fidelity. On the other hand, features of a black hole are erased within this framework--the dynamic is unitary and the degrees of freedom are conserved. Thus, the information recoverability in the fundamental picture is apparently guaranteed. Besides, the physics regarding what transpires at the Page transition remain an enigma even though QES provides a means to compute the black hole entropy.

In this Letter, we show that information recovery is applicable in much broader situations and provide one possible mechanism for the Page transition. We consider the effective dynamics that encompasses a scrambling unitary preceding the nonunitary projection of scrambled states. Releasing the constraint of the entangled-basis projections, we allow an arbitrary set of scrambled modes of cardinality $|P|$ to be projected onto a local state. The model computes the von Neumann entropy of black holes consistent with the quantum extremal surface calculation and has dual interpretations as both a holographic non-isometric mapping to the fundamental description and also the nonunitary dynamics in the effective picture. Though a substantial portion of the states are annihilated after the scrambling, the model is nonetheless free from the ailment of information loss as in the HM model and recovers information at a high fidelity. Importantly, the channels for the information flow exhibit a phase transition from the Bell-basis to the local-basis around the Page time. We showcase the Yoshida-Kitaev-like decoding strategies in a Hayden-Preskill experiment and the results from quantum computer simulations. The decoding protocols of the model have the merit of minimal quantum processor requisites and are easily testable on existing quantum computers.

\begin{figure} 
  \includegraphics[width=7.8cm]{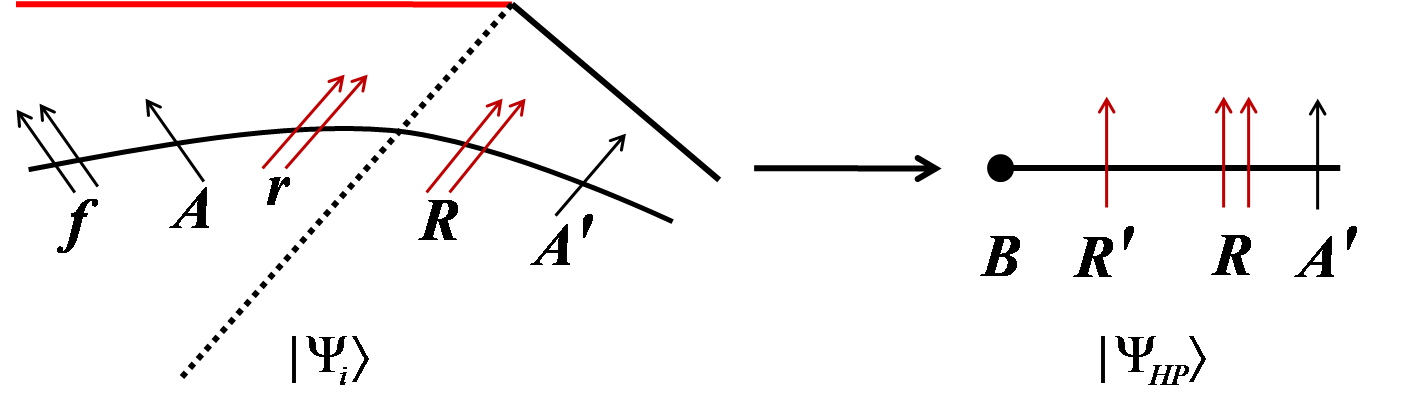}
  \caption{Left panel: Illustration of the systems in the effective field theory description on a ``nice slice". Right panel: Illustration of the systems that observed by the outside decoder.}
  \label{nice_slice_representation}
\end{figure}

{\it Nonunitary model of evaporating black holes with random local projection.--}We consider a black hole formed by a matter system $f$ in the state $|\psi_0\rangle_f$. A message system $A$, which is maximally entangled with the reference system $A'$, is thrown into the black hole at some moment. As shown in the left panel of Fig.~\ref{nice_slice_representation}, the effective field theory description on a nice slice consists of the matter system $f$, the infalling message system $A$, the reference system $A'$, as well as the outside Hawking radiation $R$ and its interior partner $r$. In this setup, the initial state of the total system is
\begin{eqnarray}\label{psi_i}
    |\Psi_{i}\rangle = |\textrm{EPR}\rangle_{A'A} \otimes |\psi_0\rangle_f \otimes |\textrm{EPR}\rangle_{rR}\;,
\end{eqnarray} 
where $\textrm{EPR}$ represents the maximally entangled state, for example, $|\textrm{EPR}\rangle_{A'A}=\frac{1}{\sqrt{|A|}}\sum_j |j,j\rangle$. 

Assuming that an external decoder is collecting the radiation and trying to recover the information stored in $AA'$, for this decoder, the degrees of freedom inside the black hole are collectively denoted by $B$. We denote the new Hawking radiation generated after $A$ falls into the black hole by $R'$, and use the random unitary operator $U$ to describe the scrambling dynamics of the black hole interior and the evaporation process. In addition, to reconcile the inconsistency between the effective and the fundamental descriptions, the degrees of freedom of the system $P$, which is randomly chosen from the outcoming states of the unitaries, are projected onto $|0\rangle_P$. Thus, the initial state of the whole system evolves into the the following modified Hayden-Preskill state 
\begin{eqnarray}\label{psi_hp}
    |\Psi_{\textrm{HP}}\rangle &=& \sqrt{|P|} \langle 0|_P \left(I_{A'}\otimes U_{(Afr)(BPR')}\otimes I_{R}\right) |\Psi_{i}\rangle\nonumber\\
    &=&\sqrt{|P|}~~~\figbox{0.17}{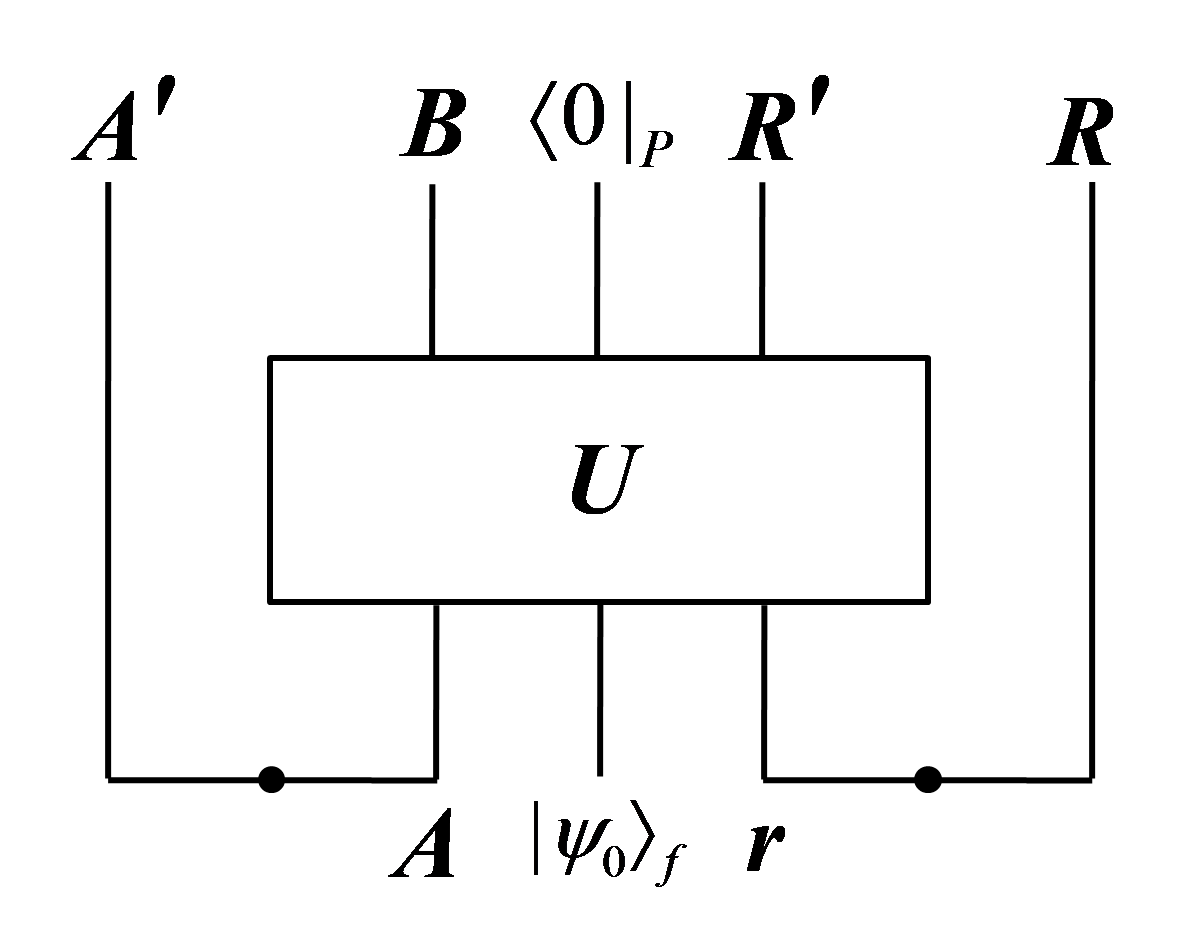}\;,
 \label{Fig:psi_hp}
\end{eqnarray}
where $\figbox{0.15}{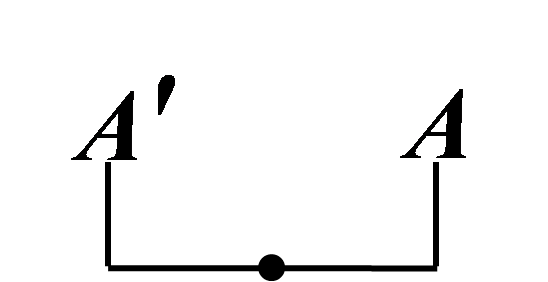}$ represents the $\textrm{EPR}$ state of $A$ and $A'$ and the black dot stands for the normalization factor $\frac{1}{\sqrt{|A|}}$. This process, when treated as the map from the effective space to the code space, is clearly a non-isometric map. Information-theoretically, the system $P$ can be selected arbitrarily from the output modes of the scrambling. The Hilbert space dimensions of the input systems $Afr$ and the output system $BPR'$ satisfy $|A||f||r|=|B||P||R'|=d$.

In this setup, $|\Psi_{\textrm{HP}}\rangle$ describes the total system in the fundamental description which includes the remnant black hole $B$, the newly-generated Hawking radiation $R'$, the early Hawking radiation $R$ and the reference system $A'$ as depicted in the right panel of Figure \ref{nice_slice_representation}. This state has the property that $\langle\Psi_{\textrm{HP}}|\Psi_{\textrm{HP}}\rangle\neq 1$  in general cases, but the normalization is preserved on the Haar average over the random unitary operator $U$, i.e., $\int dU \langle\Psi_{\textrm{HP}}|\Psi_{\textrm{HP}}\rangle=1$. In fact, though the combined operator $\sqrt{|P|} \langle 0|_P \left(I_{A'}\otimes U_{(Afr)(BPR')}\otimes I_{R}\right) $ is nonunitary, its deviation from unitarity is insignificant after averaging over the scrambling unitary. 

One further advantage of this model is that the island formula for the black hole entropy naturally emerges from it. Choosing the cutoff surface such that $B$ and $R'$ are on the inside and the rest of the exterior systems are on the outside. The $n$-th Renyi entropy is related to the density matrix by  $S_n(BR')=-\frac{1}{n-1} \log \mathrm{Tr}\left(\rho_{BR'}^n\right)$. We evaluate the Haar average of the density matrix $\rho_{BR'}^n$ and take the limit of $n=1$. The dominant contribution return the von Neumann entropy of the systems inside the cutoff surface
\begin{gather}
     S_{vN}(BR') \simeq \min \left(\log|B|+\bar S(R')\,,\bar S(A'R)\right)\,,
\end{gather}
where $\bar S$ represents the entropy in the effective description. This coincides with the quantum extremal surface calculations for black holes \cite{Engelhardt:2014gca,Penington:2019npb,Almheiri:2019psf}. The area term is represented by the coarse-grained entropy of the black hole $\log|B|$, and the entropy of the states between the cutoff surface and the quantum extremal surface is represented by $\bar S(R')$. The situation of null quantum extremal surface is given by the entropy of the systems outside the cutoff surface $\bar S(A'R)$.

{\it High-fidelity information recovery and Page transition of information channels.--} For information initially stored in $AA'$, the necessary condition for it to be retrievable from the exterior modes is that the entangled partner $A'$ is disentangled from the remnant black hole $B$. Then, the entangling system of $A'$ be switch to the external states, which allows the information to escape. This is referred to as the decoupling condition \cite{Hayden:2007cs}. It imposes the constraint on the reduced density matrix of $A'B$ to be almost separable, viz.
\begin{eqnarray}
   \int dU \|\rho_{A'B}-\frac{1}{|A'||B|} I_{A'} \otimes I_B \|_1\leq \sqrt{\frac{|f|}{|P|}} \frac{|A|}{|R'|}\;.
\label{eq:decoupling}
\end{eqnarray}
Therefore, when the new radiation satisfies $|R'| \gg \sqrt{\frac{|f|}{|P|}} |A|$, the state of $A'B$ becomes the product state and the decoupling condition is satisfied. Rewriting the decoupling condition as $|R'| \gg \frac{|B|}{|R|}|A|$, then we notice that at the early stage of the evaporation, significantly more qudits of $R'$ need to be collected in order to retrieve the information. For $|R'|,|A| \ll |f|,|R|,|B|$, few qudits of radiation in $R'$ are required when the condition $|B|\ll |R||R'|$ is satisfied a very short period after the Page time. One may notice that at late times, the degrees of freedom of the projected system satisfies $|P|>|B|$. Thus, the majority of the modes entering $U$ are randomly chosen and locally annihilated nonunitarily.

\begin{figure}
  \centering
  \includegraphics[width=.9 \columnwidth]{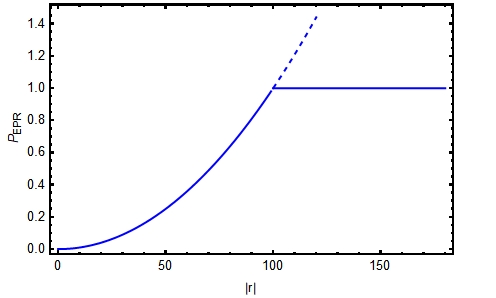}
  \caption{The EPR projection probability $P_{\mathrm{EPR}}$ as a function of the amount of early radiation $|r|$. Here, we set $|f|=10^4\,, |A|=1$. In this case, the transition is at the Page time defined by $|r|=\sqrt{|f|}=100$.}
  \label{Fig:pepr}
\end{figure}

We use the Yoshida-Kitaev protocols as an example to explicitly demonstrate how information recovery is accomplished through local projections. The computation details can be referred to Ref.~\cite{Li:2023rue}. Here, a complex conjugate of the random unitary operator $U^*$ is prepared to operate on the early radiation $R$ and two auxiliary systems $|\psi_0\rangle_f$ and $F$ imitating the systems $|\psi_0\rangle_f$ and $A$ inside the black hole. The system $F'$, which is entangled with $F$, is prepared to recover the information in $A$. The post-selected initial state, denoted as $|\Psi\rangle_{in}$, is graphically represented by  
\begin{eqnarray}
  |\Psi\rangle_{in}=C~ \figbox{0.17}{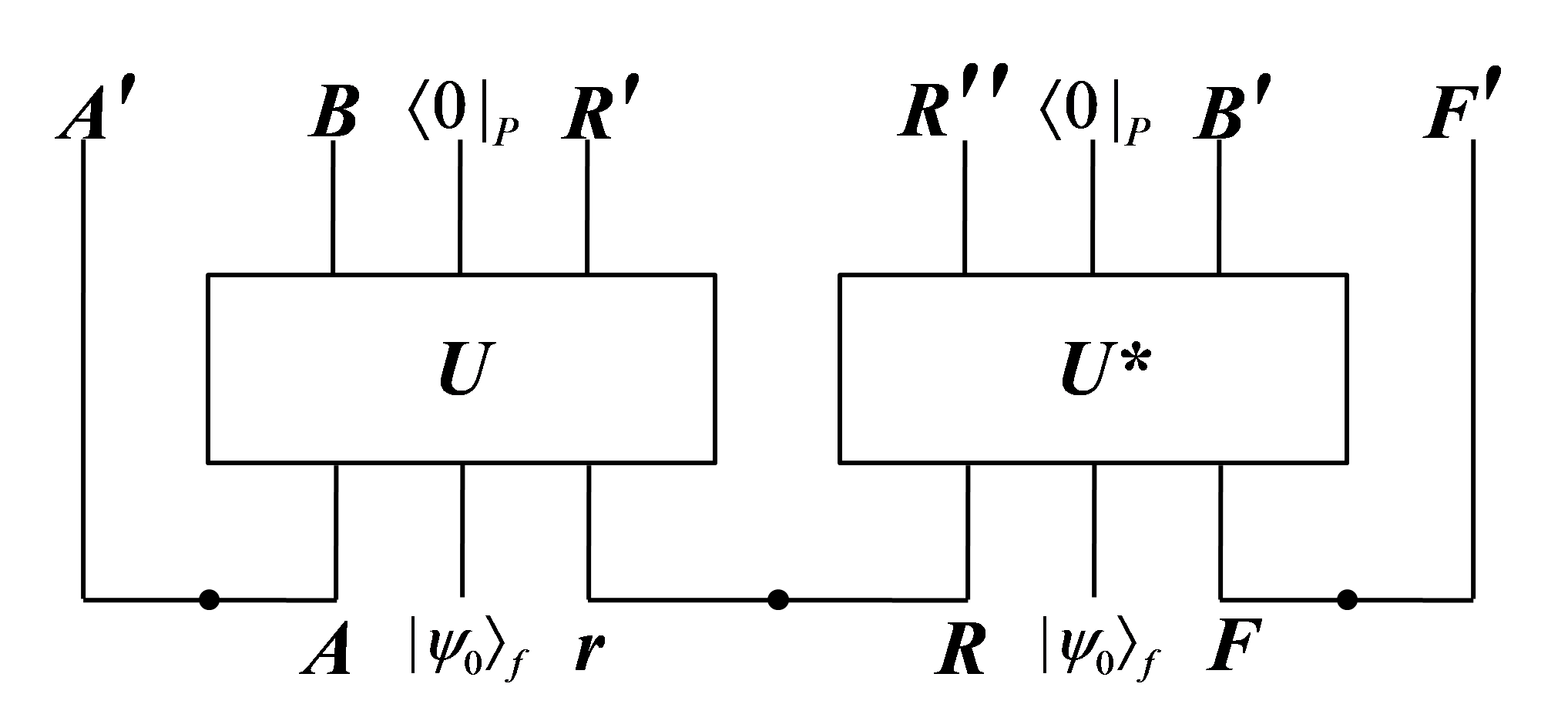}\;,
  \label{Eq:psi_in}
\end{eqnarray}
where $C=\min\left(1,\sqrt{\frac{|f|}{|P|}}|A|\right)|P|$ is the normalization factor. Typically, at the early stage of evaporation 
$C=|P|$ and at the late stage $C=\sqrt{\frac{|f|}{|P|}}|A||P|$. In fact, the non-smoothness of the function indicates certain transition of the entanglement structure around the Page time, which will be elaborated later. The operator $U^*$ can be treated as the time reversal operator of the black hole dynamics. Next, we project the systems $R'$ and $R''$ onto the EPR state $|\textrm{EPR}\rangle_{R'R''}$ by acting on it the operator $\Pi_{R'R''}=|\textrm{EPR}\rangle_{R'R''}\langle \textrm{EPR}|_{R'R''}$. The resulting state is denoted as $ |\Psi\rangle_{out}=\frac{1}{\sqrt{P_{\textrm{EPR}}}} \left(I_{A'B}\otimes\Pi_{R'R''}\otimes I_{B'F'} \right)|\Psi\rangle_{in}$, where $P_{\textrm{EPR}}$ is the probability of the EPR-projection and the factor $\frac{1}{\sqrt{P_{\textrm{EPR}}}}$ is introduced to preserve the normalization of $|\Psi\rangle_{out}$ averaged over Haar random unitaries. Under the decoupling limit, the probability of a successful information decoding is approximately
\begin{eqnarray}
     P_{\textrm{EPR}}\simeq \min\left( 1,\frac{|P|}{|f|}\frac{1}{|A|^2}\right)\,.
     \label{Eq:P_epr}
\end{eqnarray}
Imposing the information constraint $|A||f|/|r|=|B||R'|$, we can rewrite the probability of the EPR projection as $P_{\textrm{EPR}}\simeq \min\left( 1,\frac{|R|^2}{|f||A|^2}\right)$ as shown in Figure~\ref{Fig:pepr}. The non-smooth transition of the function is directly related to the transition of information channels which we will discuss in a subsequent paragraph. At the late stage when $|R|\gg |B|$, the entanglement between $R'$ and $R''$ is guaranteed and the projection becomes unnecessary. Generally, the projection $\Pi_{R'R''}$ serves to decouple the system $F'$ from $B$ and $B'$, and swap the entanglement in $AA'$ to that in $A'F'$. This can be seen by calculating the decoding fidelity, viz.
\begin{align}
    F_{\textrm{EPR}}=\textrm{Tr}\left(\Pi_{A'F'}|\Psi\rangle_{out}~_{out}\langle\Psi| \right)\simeq 1\,.
\end{align}
The high-fidelity suggests that $A'$ and $F'$ are successfully entangled and the decoding is mostly error-free. 

\begin{figure}
  \centering
  \includegraphics[width=.98\columnwidth]{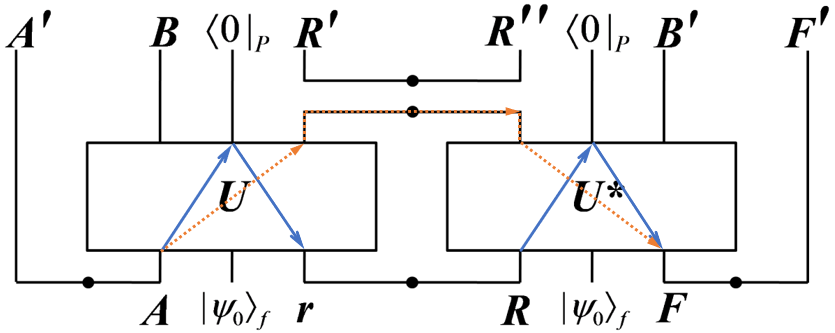}
  \caption{Phase transition of information channels from EPR projections on $|\mathrm{EPR}\rangle_{R'R''}$ (dashed orange lines) to local projections (solid blue lines). The channel for the information transmission in the case of a successful decoding is switched from the EPR projection to the local projection around the Page time. The information initially stored in the entanglement between $A$ and $A'$ is transmitted to the entanglement between $A'$ and $F'$.}
  \label{Fig:psi_out_arrow}
\end{figure}


One salient feature in the decoding process is that the size requirement for the new radiation $|R'|$ from the decoupling condition is released at late times of the evaporation. $R'$ can be as little as one qudit or even the null set. Meanwhile, the fact that the EPR projection of $R'$ and $R''$ has the probability of unity at late times ensures that the nonlocal projection can be omitted altogether. It is worth noting from $P_{\mathrm{EPR}}$ that for small message systems $|A|\ll |f|$, the transition point is approximately the Page time defined by $|f|=|R|^2$. Prior to the transition, local projections alone are insufficient to teleport the information to the exterior. Successful information recovery can only be accomplished through EPR projections. While at later stages of the evaporation post-transition, the information is transmitted through a different channel - the local projections onto $\langle 0|_P$, which create the entanglement between $A'$ and $F'$. In this scenario, information retrieval can be accomplished without any EPR projections, which was not observed in the studies of the probabilistic decoding strategies of Hawking radiations. The pictorial depiction of the two channels of the information flow are shown in Figure~\ref{Fig:psi_out_arrow}. 

To eliminate the probability associated with the EPR projection, the protocol similar to Grover's search can be employed to the post-selected state. The application of one iteration of the search operator on the state $|\Psi\rangle_{in}$ rotates this state on the two dimensional plane spanned by $|\Psi\rangle_{in}$ and $|\Psi\rangle_{out}$ by the angle $\theta$ towards the desired state $|\Psi\rangle_{out}$, where $\sin\frac{\theta}{2}=\sqrt{\frac{|P|}{|f|}}\frac{1}{|A|}$. The proof that this protocol satisfies the requirements of the decoding is presented in Ref.~\cite{Li:2023rue}. After $n$ times of iterations, the outcoming state becomes the superposition of the desired state with its perpendicular state,
\begin{eqnarray}
|\Psi(n)\rangle=\sin\left((n+1/2)\theta\right)|\Psi\rangle_{out}+\cos\left((n+1/2)\theta\right)|\Psi\rangle_{out}^{\perp}\;.
\end{eqnarray}


{\it Quantum circuit representations and experiments on quantum computers.--}The information recoverability of the model can be tested on 7-qubit quantum processors \cite{remark1}. For simplicity, we assume a pure-state message system $A$ without the reference $A'$ and release the constraints on the dimensions of the systems. The explicit circuits are given in Ref.~\cite{Li:2023rue}, where the circuit of the scrambling unitary is chosen such that it transforms all single-qubit operations into three-qubit operations.

\begin{figure}
  \centering
    \includegraphics[width=.49\columnwidth]{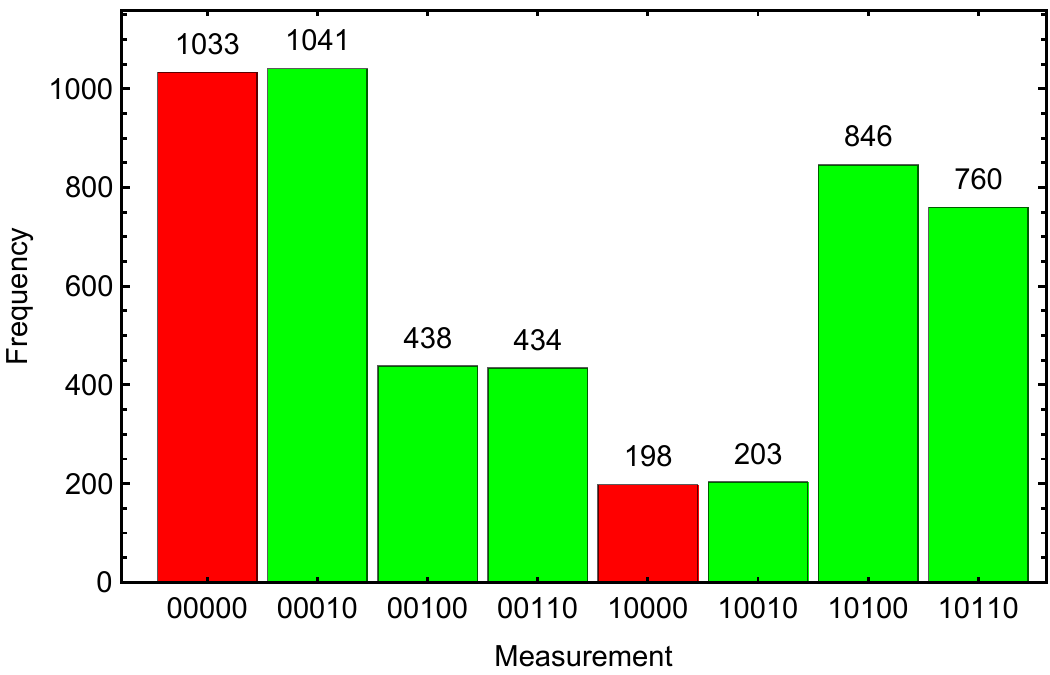}
    \includegraphics[width=.49\columnwidth]{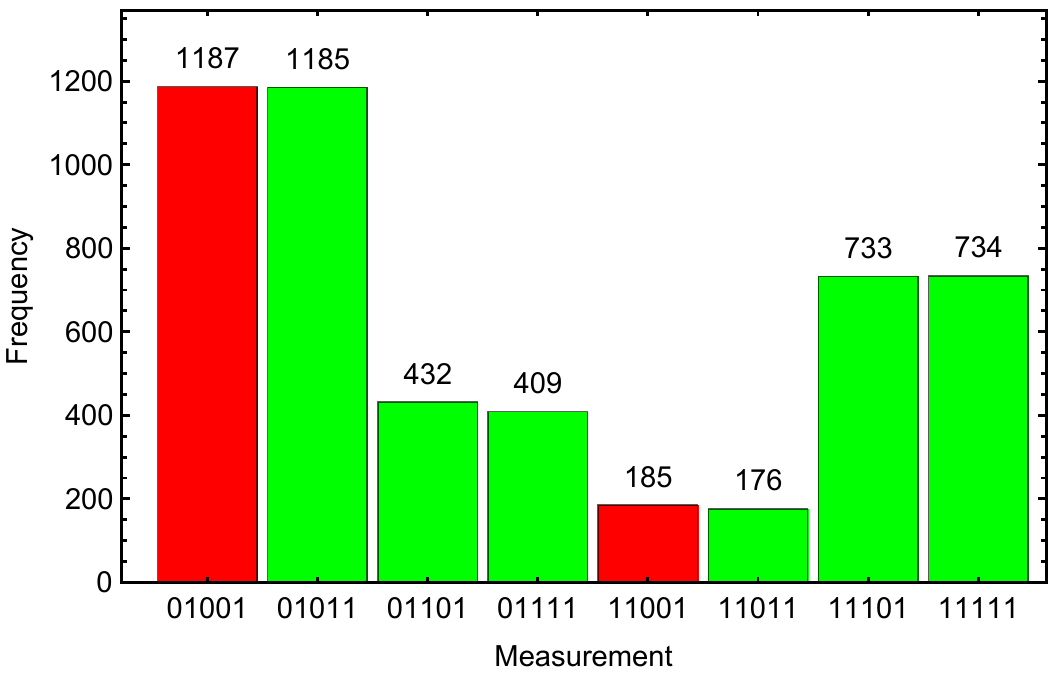}
    \includegraphics[width=.49\columnwidth]{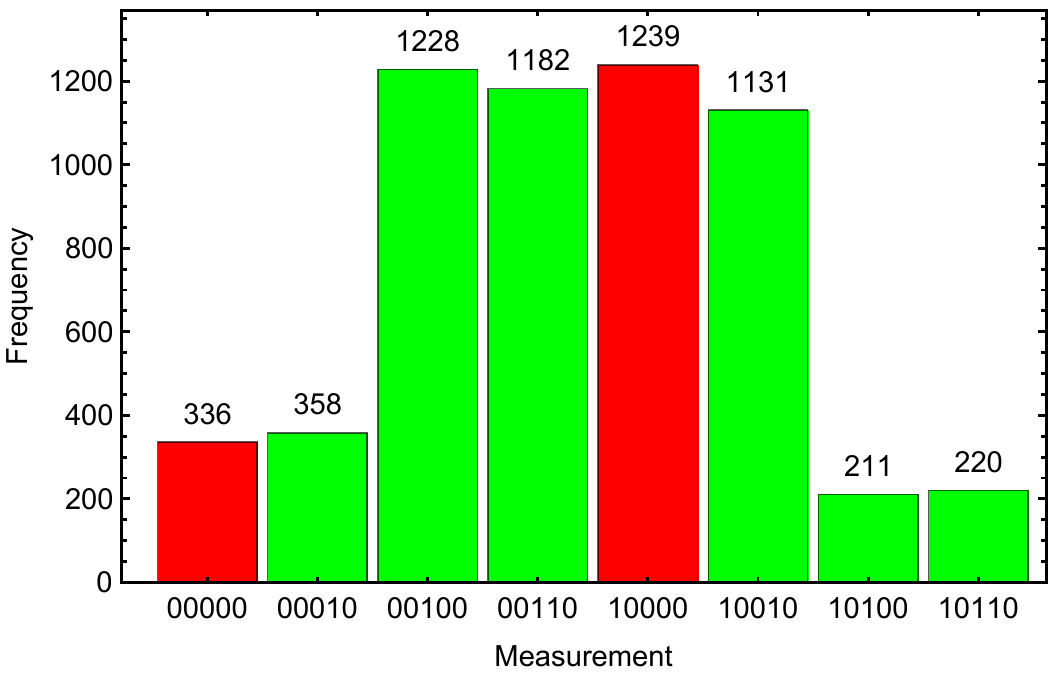}
    \includegraphics[width=.49\columnwidth]{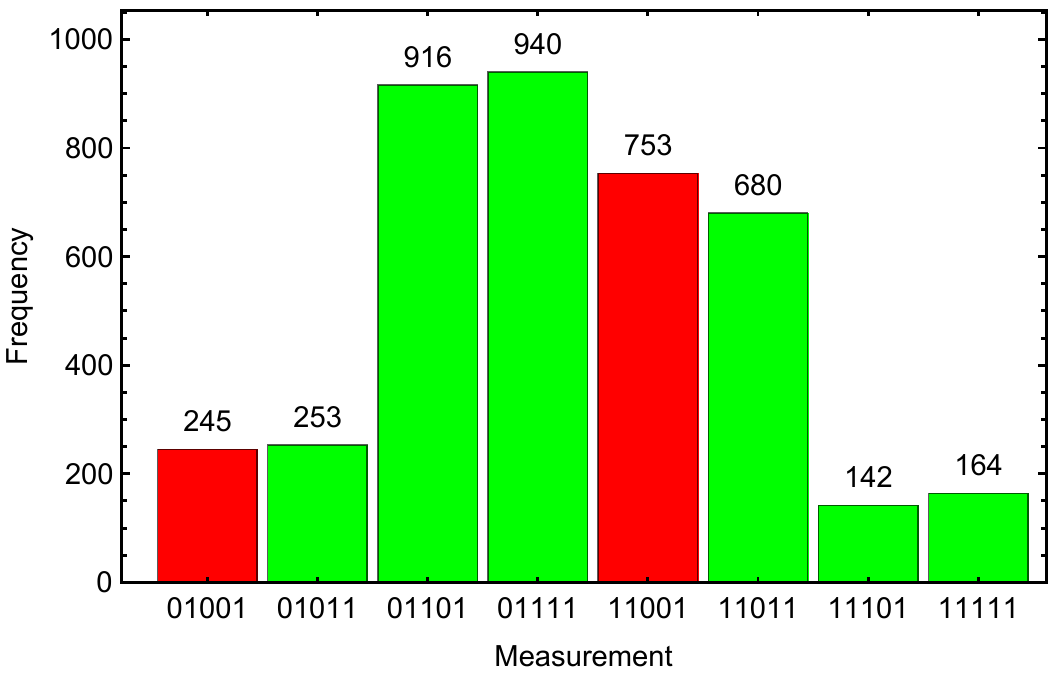}
\caption{Experimental results of the decoding through EPR projections on IBM-nairobi for 20,000 shots. The horizontal axis represents the measurement results in the form of $\textrm{q}[6]\textrm{q}[4]\textrm{q}[3]\textrm{q}[2]\textrm{q}[1]$ and the vertical axis represents the corresponding frequencies. Top left panel: the initial input is q[0]=$|0\rangle$ and the qubits $\textrm{q}[4]\textrm{q}[1]$ are projected to the state $|00\rangle$. Top right panel: the initial input is q[0]=$|0\rangle$ and $\textrm{q}[4]\textrm{q}[1]$ are projected to the state $|11\rangle$. Bottom left panel: the initial input is q[0]=$|1\rangle$, and $\textrm{q}[4]\textrm{q}[1]$ are projected to the state $|00\rangle$. Bottom right panel: the initial input is q[0]=$|1\rangle$ and $\textrm{q}[4]\textrm{q}[1]$ are projected to $|11\rangle$.} 
  \label{q0=0barchart}
\end{figure}

The decoding strategy based on EPR projections is implemented on the IBM-nairobi quantum processor as shown in Figure~\ref{q0=0barchart}. The qubits q[0] to q[6] represent $A$, $f$, $r$, $R$, $f$, $F$, and $F'$, respectively. The red bars represent the correct projection of $R'$ and $R''$ onto the $\textrm{EPR}$ state $\frac{1}{\sqrt{2}}\left(|00\rangle+|11\rangle\right)$. The green bars represent incorrect projections to other $\textrm{EPR}$ states, which indicate a failed decoding. From Figure~\ref{q0=0barchart}, the probability of projecting to the correct $\textrm{EPR}$ state $\frac{1}{\sqrt{2}}\left(|00\rangle+|11\rangle\right)$ is around $25\%$ to $27\%$. Theoretically, projections on the correct $\textrm{EPR}$ state indicate a success in decoding the radiation. In reality, there are errors due to noises in the quantum computers. The decoding efficiency, defined as the ratio between the frequency of a successful decoding to the frequency of a successful $\textrm{EPR}$ projection, is about $79\%$ to $87\%$ in the four cases. As an example, the red bar with q[6] = $|0\rangle$ in Fig.~\ref{q0=0barchart}(a) has a much higher value than the red bar with q[6] =$|1\rangle$, which suggests a high probability of decoding the initial information q[0] = $|0\rangle$.

\begin{figure}[tb]
  \centering
  \includegraphics[width=.49\columnwidth]{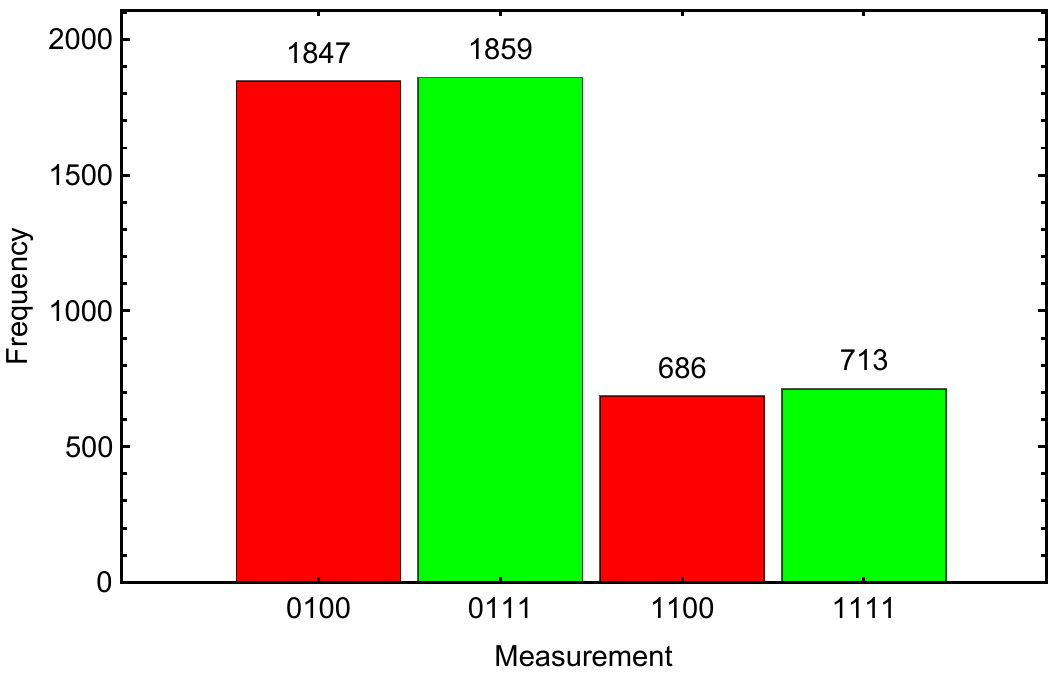}
    \includegraphics[width=.49\columnwidth]{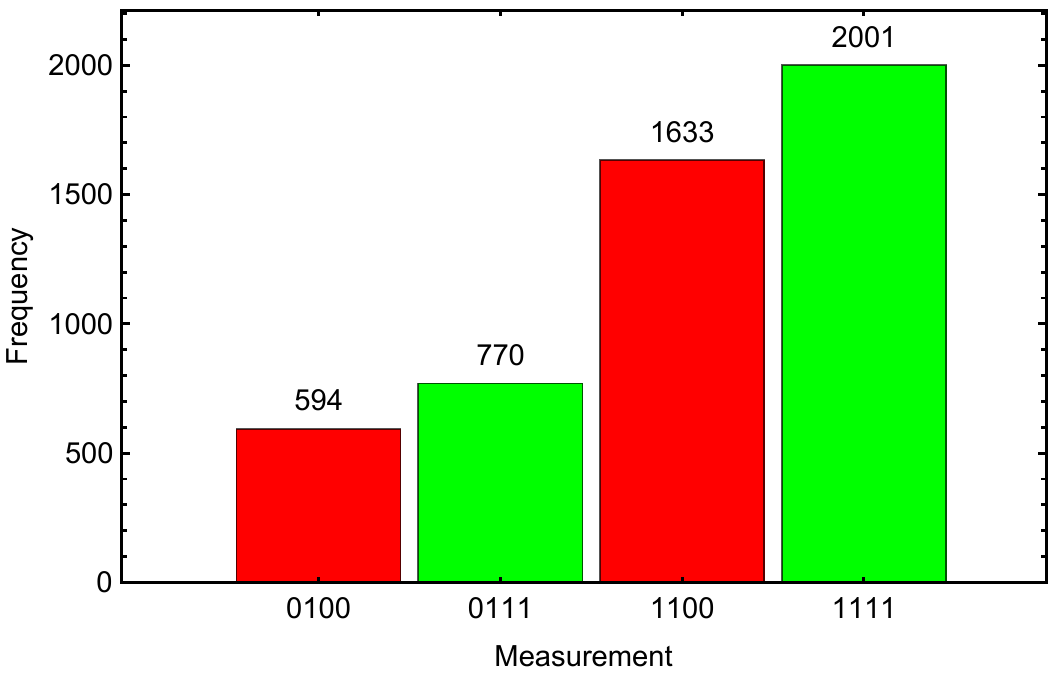}
  \caption{Experimental results of Grover's search decoding algorithm on the IBM-perth for 20,000 shots. Left: the statistical results for the initial input of $\textrm{q}[0]=|0\rangle$. Right: the statistical results for the initial input of $\textrm{q}[0]=|1\rangle$.}
  \label{Grover_q0=0}
\end{figure}

We performed the tests of the Grover's search decoding strategy on the IBM-perth processor. The qubits q[0] to q[5] represent $A\,,f$, $r$, $R\,,f$ and $F$, respectively. Since the systems $A$, $f$ and $P$ are all represented by one qubit, $|P|=|f|=|A|=2$ and the reflection angle $\theta=\frac{\pi}{3}$. Theoretically, the initial quantum state of $A$ is accurately recovered by applying the operators on the state $|\Psi\rangle_{in}$ only one time. The experimental data are shown in Figure~\ref{Grover_q0=0} where the red bars represent post-selecting the two $P-$projections to be $|00\rangle$ and the green bars are for $|11\rangle$. For both the initial input q$[0]=|0\rangle$ and q$[0]=|1\rangle$, the decoding efficiencies are approximately $72\%$ to $73\%$. We notice that the decoding efficiencies are slightly compromised compared to the probabilistic decoding strategy due to the higher circuit complexity. However, there is no additional probability of EPR projections on which the eventual decoding efficiencies are conditioned. Therefore, the overall decoding efficiencies of the Grover's search decoding strategy are higher.

{\it Conclusion.--}
Decoding Hawking radiation has been explored primarily within the framework of a unitary black hole dynamics. However, a crucial question arises when a substantial reduction in degrees of freedom occurs during the transition from the effective to the fundamental descriptions. In contrast to the HM model, the interior modes are scrambled before the local projections. This self-averaging behavior of the unitary interactions in the bulk is supported by studies of the JT gravity and the SKY model \cite{Saad:2018bqo,Saad:2019lba}. Unlike the dynamical non-isometric code models, the resulting states from the projection need not be maximally entangled with the right-moving modes in the interior. Information-theoretically, projections onto maximally-entangled states facilitate a direct transfer of information by connecting the interior and external right-moving states. By locally projecting the annihilated states following the unitary scrambling, the channel for information transmission transitions from the EPR projections to the local projections, reminiscent of phase transitions induced by local measurements. The information transmission through this channel becomes more faithful when a larger portion of the modes entering the scrambling unitary are projected.

In conclusion, our study establishes the viability of information retrieval from the black hole interior within a nonunitary framework involving the local annihilation of random states. The island formula for black hole entropy is obtained and the high-fidelity information recoveries through probabilistic and Grover's search strategies are realized. We show that the two channels of information transmission--through EPR projections and local projections--feature a phase transition-like behavior around the Page time. This offers a new perspective on the Page transition. Furthermore, we implement decoding protocols through the quantum circuits and conduct tests of the strategies on the IBM quantum computers using a full scrambling unitary circuit. Partial scrambling unitaries designed for specific qubit configurations on IBM quantum processors have been widely used \cite{Yan:2020fxu,Harris:2021mma}. The issue with such unitaries is that they do not satisfy the scrambling properties given in \cite{Landsman:2018jpm,Li:2023rue} such that decoding information from those partial-scrambling unitaries frequently fails due to the information loss in this model. The simulation of the quantum circuits on IBM quantum processors verifies the viability of information retrieval through two channels, and demonstrates the practicality of high-quality three-qubit scrambling in two decoding strategies.

{\it Acknowledgments.--}We acknowledge the IBM Quantum services for this work. X.W appreciates the start-up grant from the Wenzhou Institute of UCAS. 

*X.~Wang and R.~Li contribute equally to this work.

\end{document}